\documentclass{PoS}

\usepackage{xspace}
\usepackage{upgreek}
\usepackage[numbers]{natbib}
\usepackage[percent]{overpic}
\usepackage{color}

\newcommand{\suz}{\textit{Suzaku}\xspace}
\newcommand{\hess}{H.E.S.S.\xspace}

\title{The Vela~X pulsar wind nebula through the eyes of \hess and \suz}

\ShortTitle{Vela~X with \hess and \suz}

\author{\speaker{L.~Tibaldo}$^a$, F.~Aharonian$^{a,b,c}$, P.~Bordas$^a$, S. Caroff$^d$, J.~A.~Hinton$^a$, D.~Khangulyan$^e$, H.~Odaka$^f$, R.~Tuffs$^a$, for the \hess Collaboration\\
E-mail: \email{luigi.tibaldo@mpi-hd.mpg.de}

{\footnotesize
        $^a$ Max-Planck-Institut f\"ur Kernphysik, P.O. Box 103980, D 69029 Heidelberg, Germany\\
        $^b$ Dublin Institute for Advanced Studies, 31 Fitzwilliam Place, Dublin 2, Ireland\\
        $^c$ National Academy of Sciences of the Republic of Armenia, Marshall Baghramian Avenue, 24, 0019 Yerevan, Republic of Armenia\\
        $^d$ Laboratoire Leprince-Ringuet, Ecole Polytechnique, CNRS/IN2P3, F-91128 Palaiseau, France\\
        $^e$ Department of Physics, Rikkyo University, 3-34-1 Nishi-Ikebukuro, Toshima-ku, Tokyo 171-8501, Japan\\
        $^f$ Japan Aerpspace Exploration Agency (JAXA), Institute of Space and Astronautical Science (ISAS), 3-1-1 Yoshinodai, Chuo-ku, Sagamihara, Kanagawa 229-8510, Japan\\
	}}

\abstract{Pulsar wind nebulae (PWNe) are among the most extreme particle accelerators in galaxies, are recognized as multi-TeV electron/positron sources, and are one of the dominant classes of Galactic gamma-ray sources. Vela~X is a nearby PWN at 290~pc from the Earth with large apparent size ($>1^\circ$). The \hess array of imaging atmospheric Cherenkov telescopes has detected Vela~X as one of the brightest known sources of TeV gamma rays. The bulk of the gamma-ray emission measured using \hess coincides with an elongated structure known from X-ray observations and dubbed the cocoon, that seemingly emanates from the region of the pulsar wind termination shock. The spectral energy distribution of the cocoon peaks at around 10~TeV, and then presents a cutoff that can be precisely measured with \hess owing to the extreme brightness of the source. Electrons radiating inverse-Compton gamma rays in the cutoff region are the same responsible for the X-ray synchrotron emission at energies $> 1$~keV. Therefore, Vela~X provides a unique test case, in which we can constrain the densities and spectra of accelerated leptons in the cutoff regime, as well as the magnetic field properties, with minimal modeling assumptions. Thanks to the proximity/large apparent size of the source, this can be done in a spatially-resolved fashion across the PWN. We will present an analysis of \hess data combined with X-ray data from the \suz space telescope. We will discuss implications for the mechanisms behind particle acceleration and transport, constrain the strength of the magnetic field in different locations in the nebula, and probe for magnetic field turbulence.}

\FullConference{35th International Cosmic Ray Conference --- ICRC2017\\
		10--20 July, 2017\\
		Bexco, Busan, Korea}

\begin{document}

\section{Introduction}

Vela~X is one of the archetypal TeV pulsar wind nebulae (PWNe), i.e., a system in which relativistic particles from a pulsar wind emit gamma rays up to the highest observable energies. Owing to the proximity of this system (290 pc), and, thus, large apparent size ($> 1^\circ$), and to its extreme brightness, Vela~X is well resolved by gamma-ray atmospheric Cherenkov Telescopes. TeV gamma-ray emission is largely correlated with an elongated structure seen in X-rays dubbed the ``cocoon'', that extends from the pulsar wind termination shock for $\gtrsim 1^\circ$ (5~pc) \cite{hessvelax2012}.

Hydrodynamical simulations \cite{blondin2001} suggest that the cocoon might have formed as a result of the supernova remnant reverse shock impact on the PWN. The cocoon spectrum measured by the \hess telescopes extends beyond 60~TeV \cite{hessvelax2012}, tracing extremely energetic particles (electrons with energies up to $>100$~TeV). On the other hand, the typical magnetic field values of order $\sim$5~$\upmu$G, suggested by X-ray observations, correspond to 100~TeV electron cooling times due to synchrotron energy losses of 4~kyr, much shorter than the age of the system of 20~kyr inferred from the hydrodynamical simulations. This suggests that efficient particle acceleration and/or transport is at play in the cocoon, the nature of which still remains an open question. 

In this paper we present a joint analysis of three regions across the cocoon using X-ray data from the \suz space telescope and \hess Synchrotron emission measured by \suz and inverse-Compton emission measured by \hess trace the same parent electron population with energies from 30~TeV to $>100$~TeV, where \hess data are known to indicate a cutoff in the particle spectrum. By combining \suz and \hess data we derive the shape of the electron spectrum cutoff, which encodes information on the acceleration and propagation mechanisms, and the properties of the magnetic fields in a spatially resolved fashion across the cocoon with minimal modeling assumptions.  

\section{Observations and Definition of Regions for Spectral Analysis}

\begin{figure}
     \begin{center}
     \begin{overpic}[width=1.\textwidth]{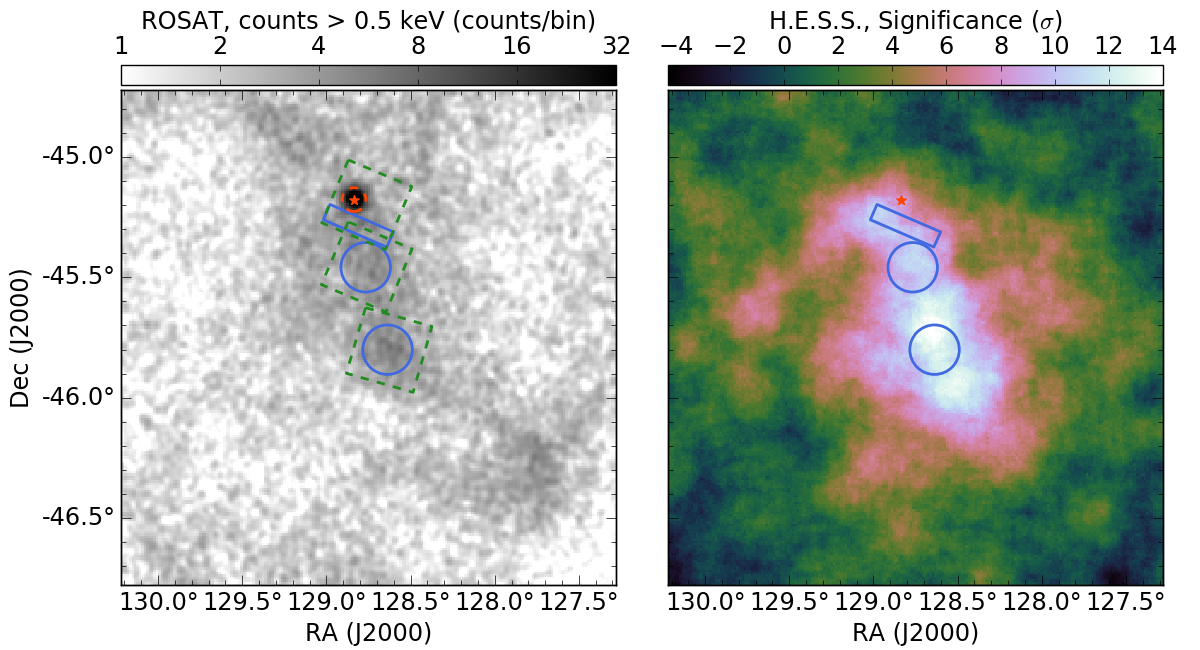}
     \put(58,7){\color{white}H.E.S.S. 2017 Preliminary}
     \end{overpic}
     \caption{Regions used for spectral analysis (blue box and circles) are overlaid to two maps of the Vela~X region. On the left: X-ray count map from the \textit{ROSAT} survey at energies $>0.5$~keV. On the right: significance map from \hess at energies $>0.6$~TeV (see text for details on how the map is derived). The red star indicates the position of the Vela pulsar and its size approximately corresponds to that of the jet-torus structure measured with \textit{Chandra} \cite{manzali2007}. Also shown in the left image: borders of the \suz field of view for the three pointings used in this paper (dashed green squares); 95\% containment radius of the \suz point spread function around the Vela pulsar (dashed red circle).}
     \label{fig:maps}
     \end{center}
     \vspace{-30pt}
     \end{figure}

Owing to its field of view of 18~arcmin and good energy coverage up to 10 keV, the \suz X-ray Imaging Spectrometer (XIS) provides the best existing dataset to study the extended X-ray emission from electrons in the cutoff region from the Vela X cocoon. We use data from three observations performed with \suz in 2006. The three observations cover different regions across the cocoon, as illustrated in Fig.~\ref{fig:maps}, i.e., from North to South: 0) toward the Vela pulsar and inner PWN (60~ks), 1) and 2) toward the cocoon (61~ks and 18 ks, respectively).
We combine \suz observations with gamma-ray data accumulated with \hess from 2004 to 2016 as described below, for a total of $>100$~h of livetime.

For each \suz pointing we derive the X-ray and gamma-ray spectral energy distributions (SEDs) from the very same extraction region, so that we can subsequently derive the electron and magnetic field properties. For pointing~1 and 2 we use a circular spectrum extraction region with a radius of 7.5 arcmin centred on the \suz boresight. For pointing 0 we select a smaller box extraction region at minimal distance from the pulsar of 3.6~arcmin (95\% containment radius of the XIS point spread function, equivalent to 0.3~pc at the source distance) to exclude X-ray emission produced by thermal emission from the surface of the neutron star as well as by nonthermal magnetospheric processes. 
This also excludes from the analysis the innermost part of the PWN that shows a bright jet-torus structure in X rays \cite{manzali2007} not resolved in TeV gamma rays. The spectrum extraction regions are shown in Fig.~\ref{fig:maps}.

\section{Analysis}

We process the X-ray data using the \suz software and applying standard event screening criteria. We restrict the analysis to energies $>2.25$~keV to exclude thermal emission from the Vela supernova remnant. We note that contamination from the Galactic ridge X-ray emission is negligible in our analysis regions. We evaluate the instrumental background based on night Earth's observations using the \texttt{xisnxbgen} tool. The spectra in all regions are well fitted with a single power law modified by a fixed Galactic interstellar absorption with hydrogen column density $N_\mathrm{H}=2.59 \times 10^{20}$~cm$^{-2}$ \cite{manzali2007}, on top of the cosmic X-ray background parametrised  as in \cite{miyaji1998}. Based on this spectral analysis, we derive the X-ray SED in 10~energy bins between 2.25~keV and 9.5~keV. We add in quadrature to the statistical uncertainties a 10\% systematic uncertainty on flux measurements \cite{sekiya2016}. Furthermore, we add to the uncertainty in the SED points of pointing~0 the estimated spill-over into the analysis region from the region within 3.6~arcmin from the pulsar position, estimated assuming that all emission comes from a pointlike source at the position of the pulsar.

For this analysis focused on high energies, we use only gamma-ray data from the four 12~m diameter \hess telescopes, selecting observing periods where at least three out of four of these telescopes were operational. Event reconstruction and analysis is based on a multivariate technique applied to the atmospheric shower image shape parameters \cite{ohm2009}. All the results have been cross-checked with an independent calibration and analysis chain based on an air-shower model template approach \cite{denaurois2009}.  In both analysis chains we use the most stringent selection criteria to exclude background from cosmic-ray showers. Furthermore, to ensure uniform energy thresholds among different observing times, thus reducing systematics, we include only candidate gamma-ray events with reconstructed energy $> 0.6$~TeV in the subsequent analysis. 

The residual background from cosmic-ray showers is evaluated from data taken within the same pointing, excluding regions of the field of view that display significant gamma-ray emission according to the results of the \hess Galactic Plane Survey, or where the brightness temperature at 44~GHz measured by Planck is $> 1.5$~mK (indicating the presence of relativistic electron populations that may radiate in gamma rays). The significance image in Fig.~\ref{fig:maps}~(right) is estimated using the ring background method \cite{berge2007}, and subsequently applying the formula by Li~\&~Ma \cite{lima1983}.

For the extraction of the spectra from  the three analysis regions we apply the reflected background method \cite{berge2007}. The requirement to have at least two reflected regions for background estimation outside the exclusion region reduces the livetime to 70~h, 75~h, and 80~h for pointings~0, 1, and 2 respectively. We fit to the resulting count spectra power laws and power laws with exponential cutoffs. The latter result statistically preferred in pointings~1 and 2. For pointing~0 the best-fit exponential-cutoff power law has a cutoff energy consistent within statistical uncertainties with the other two regions, but the presence of a cutoff is not statistically significant ($<3\sigma$).
Using the best-fit spectral function we derive a binned SED for each pointing, requiring that each SED point has a minimum statistical significance of $2\sigma$. Statistical uncertainties in the SED points are combined in quadrature with systematic uncertainties of 40\%, 31\%, and 35\% for pointings~0, 1, and 2 respectively, which result from adding 20\% systematic uncertainties on the flux measurements \cite{aharonian2006crab} to the systematic differences found between the  main and alternative analyses. 

\section{Multiwavelength Spectral Fitting with Radiative Models}

We model the X-ray and gamma-ray SEDs with a simple radiative model that assumes a single population of relativistic electrons for each region with spectrum described by the formula 
\begin{equation}\label{eq:espec}
\frac{\mathrm{d}N}{\mathrm{d}E} = A \left(\frac{E}{E_0}\right)^{-\alpha} \exp{\left[-\left(\frac{E}{E_\mathrm{co}}\right)^\beta\right]}
\end{equation}
where $N$ is electron number, $E$ is electron energy, $A$ is an electron number density  that normalises the distribution, $\alpha$ is the spectral index of a power-law distribution with reference energy $E_0$, multiplied by an exponential cutoff with cutoff energy $E_\mathrm{co}$ and cutoff index $\beta$.

The electrons produce X-ray emission through synchrotron radiation in a magnetic field,
and, at the same time, gamma-ray emission through inverse-Compton scattering on low-energy photons from the Cosmic Microwave Background (CMB), and the diffuse infrared radiation from interstellar dust. For the latter we adopt the model by \cite{popescu2017} at the position of Vela~X.  We use the \texttt{naima} software package \cite{zabalza2015} to calculate the model SEDs and perform a Markov Chain Monte Carlo (MCMC) fit of the electron spectrum and a uniform magnetic field strength to the multiwavelength data.

\begin{figure}
     \begin{center}
     \begin{tabular}{cc}
  \hspace{-1.cm}  \begin{overpic}[width=0.55\textwidth]{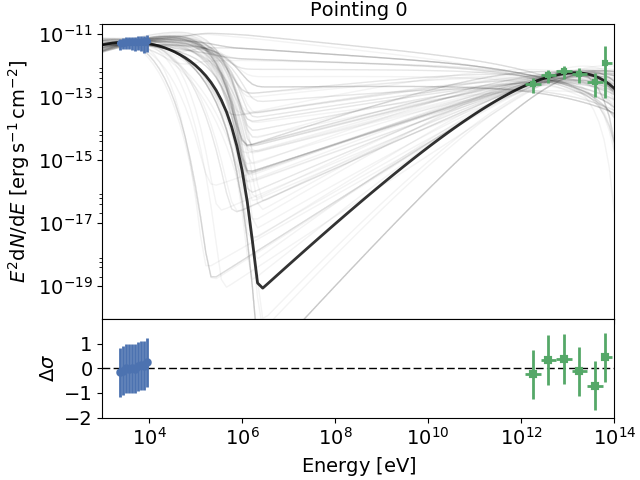}
  \put(53,28){\footnotesize H.E.S.S. 2017 Preliminary}
  \end{overpic}
  &
  \hspace{-0.5cm}   \begin{overpic}[width=0.55\textwidth]{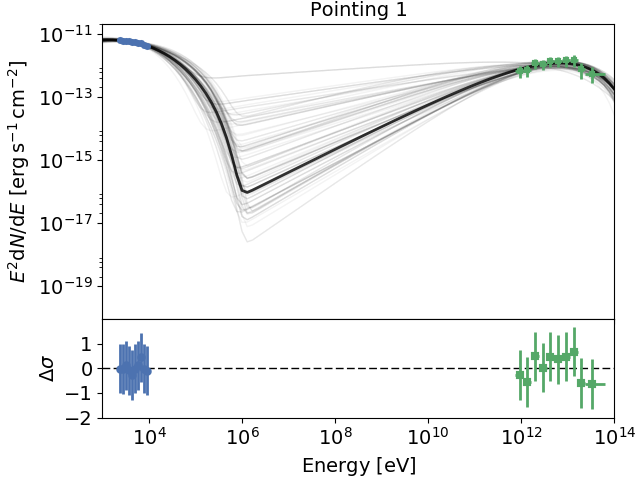}
  \put(53,28){\footnotesize H.E.S.S. 2017 Preliminary}
  \end{overpic}
  \\
   \hspace{-1.cm}   \begin{overpic}[width=0.55\textwidth]{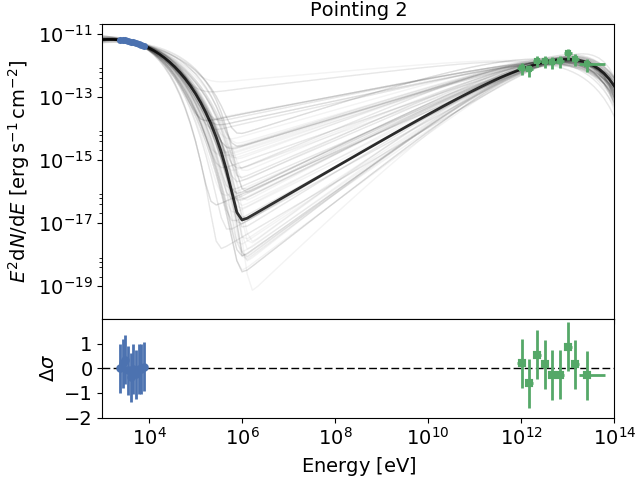}
   \put(53,28){\footnotesize H.E.S.S. 2017 Preliminary}
  \end{overpic}
    \end{tabular}
    \caption{Multiwavelength spectral energy distribution of the emission from the three regions along the Vela~X cocoon. Blue points show X-ray measurements with \suz, green points gamma-ray measurements with \hess. Error bars include systematic uncertainties (see text for details). The lines show radiative models: the grey lines show 100 random realisations from the parameter space scan, thus illustrate the model uncertainty, and the black line the best-fit model. The bottom sub-panel for each pointing shows deviations of the points from the best-fit model.}\label{fig:sed}
         \end{center}
     \vspace{-24pt}
     \end{figure}

\begin{figure}
     \begin{center}
     \begin{tabular}{cc}
    \begin{overpic}[width=0.45\textwidth]{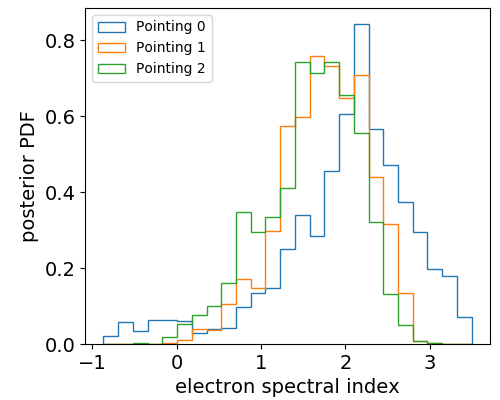}
    \put(48,81){\footnotesize H.E.S.S. 2017 Preliminary}
    \end{overpic}
    &
    \begin{overpic}[width=0.45\textwidth]{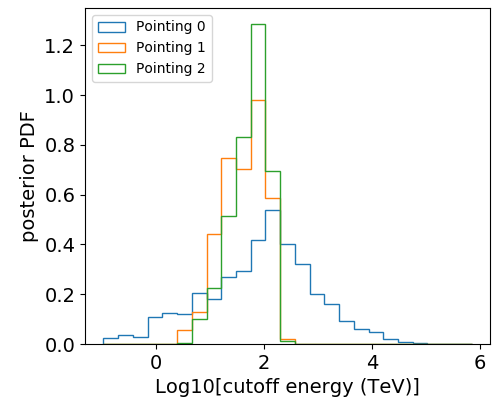}
    \put(48,81){\footnotesize H.E.S.S. 2017 Preliminary}
    \end{overpic}
    \\
    \\
    \begin{overpic}[width=0.45\textwidth]{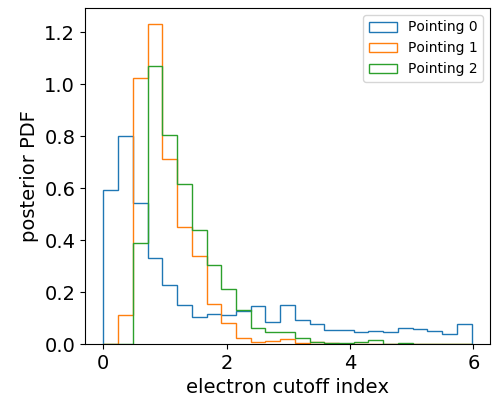}
    \put(48,81){\footnotesize H.E.S.S. 2017 Preliminary}
    \end{overpic}
    &
    \begin{overpic}[width=0.45\textwidth]{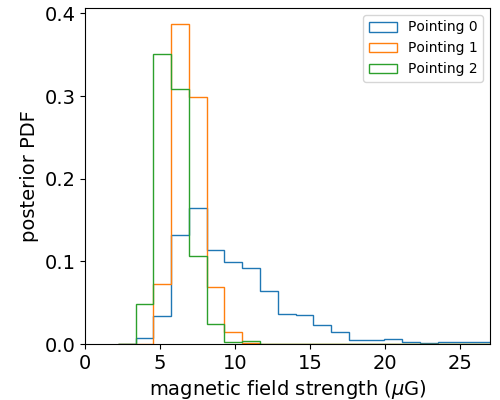}
    \put(48,81){\footnotesize H.E.S.S. 2017 Preliminary}
    \end{overpic}
    \end{tabular}
    \caption{Posterior Probability Density Function (PDF) of the parameters describing the electron spectrum (Eq.~\ref{eq:espec}) and uniform magnetic field strength for the three pointings from the joint analysis of the \suz and \hess spectral energy distributions.}\label{fig:pars-def}
         \end{center}
         \vspace{-12pt}
     \end{figure}

Fig.~\ref{fig:sed} shows the three multiwavelength SEDs along with the models from the MCMC scans. Our simple leptonic model can naturally reproduce the data in all regions.
Fig.~\ref{fig:pars-def} shows the model parameters posterior probability density functions (PDFs). The properties of the particle populations are remarkably uniform across the Vela~X cocoon from the region immediately beyond the pulsar wind termination shock to a distance of 5 pc. 
The index of the electron cutoff is not well constrained by the observations. The magnetic field strengths are $9 \pm 3$~$\upmu$G, $6.8 \pm 0.9$~$\upmu$G, and $5.8 \pm 1.0$~$\upmu$G, for pointings~0, 1, and 2 respectively.

Consistently with \cite{mangano2005}, for regions~0 there are indications of a harder particle spectrum from the X-ray observations. This could be explained by a second particle population linked to the innermost region of the PWN, overlapping in the region from pointing 0 with the cocoon population. An alternative hypothesis that can be considered is the presence of turbulence in the magnetic fields. Magnetic turbulence would harden the X-ray synchrotron spectrum \cite{kelner2013}. Magnetic turbulence is also expected to be present in conjunction with magnetic reconnection, that is a potential mechanism for fast particle acceleration inside the PWN (and has also been invoked to explain the flaring episodes of the other archetypal TeV PWN, the Crab nebula).

\begin{figure}
     \begin{center}
     \begin{tabular}{cc}
    \begin{overpic}[width=0.45\textwidth]{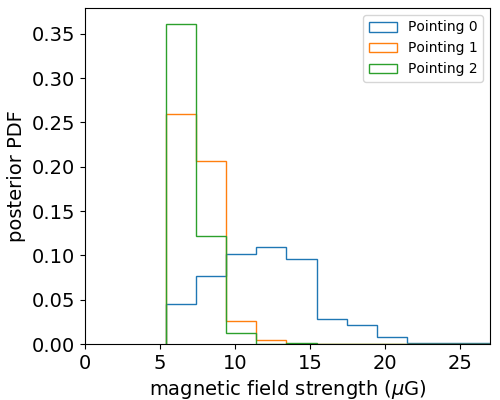}
        \put(48,81){\footnotesize H.E.S.S. 2017 Preliminary}
    \end{overpic}
    &
    \begin{overpic}[width=0.45\textwidth]{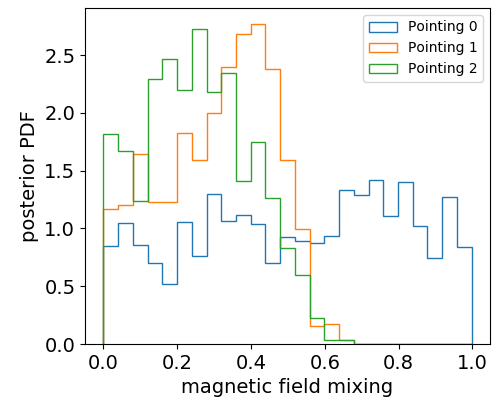}
        \put(48,81){\footnotesize H.E.S.S. 2017 Preliminary}
    \end{overpic}
    \\
    \end{tabular}
    \caption{Posterior Probability Density Function (PDF) of the parameters describing the root-mean-square magnetic field strength and the mixing parameter between uniform and turbulent magnetic field components (Eq.~\ref{eq:turbB}) for the three pointings from the joint analysis of the \suz and \hess spectral energy distributions.}\label{fig:pars-turb}
         \end{center}
         \vspace{-24pt}
     \end{figure}

We test the presence of magnetic turbulence against our data by performing a second MCMC scan in which we assume that synchrotron emission is produced on a two-component magnetic field with PDF:
\begin{equation}\label{eq:turbB}
\mathrm{PDF}(B) = (1-a) \delta(B-B_\mathrm{RMS}) + a C B^{-\alpha} H(B-B_\mathrm{min}) H(B_\mathrm{max} - B).
\end{equation} 
The parameter $a$ (between~0 and~1) defines the mixing of the two components, i.e., the turbulence level. The first component has uniform strength $B_\mathrm{RMS}$. The second component has a power-law strength distribution with index $\alpha$ between a minimum strength $B_\mathrm{min}$ and a maximum strength $B_\mathrm{max}$. We take as an example $\alpha = 1.5$ (Kraichnan spectrum), and we set $B_\mathrm{max} = 100 \times B_\mathrm{RMS}$. The parameters $C$ and $B_\mathrm{min}$ are determined so that the PDF is normalised to~1, and the root-mean-square (RMS) expectation value of the magnetic field strength for the turbulent component is $B_\mathrm{RMS}$ (since the synchrotron power is proportional to $\langle B^2\rangle$). The only remaining free parameters are thus the mixing $a$ and $B_\mathrm{RMS}$ that are fit to the data along with the electron spectrum parameters as defined above.
Fig.~ \ref{fig:pars-turb} shows that posterior PDF for these parameters for the three pointings. In pointing~0 the data cannot distinguish between the hypotheses of uniform or turbulent magnetic field. On the other hand, for pointing~1 and 2 we can constrain the level of turbulence to $<57\%$ and $<56\%$ (99\% higher percentile of the posterior PDF), respectively, for the model considered.
\vspace{4pt}

\section{Final Remarks}
A simple model with a single population of electrons producing X-ray synchrotron emission and inverse-Compton gamma-ray emission provides a very natural explanation for the multiwavelength observations of the Vela~X cocoon. The electron spectrum and strength of the magnetic field appear to be remarkably uniform across the cocoon, from distances $\sim$0.3~pc to $>5$~pc from the pulsar wind termination shock. The spectral uniformity, combined with magnetic field strengths $> 5$~$\upmu$G,  reinforces the need for an efficient mechanism to accelerate particles in the cocoon and/or to transport them from the region of the termination shock across the cocoon on timescales well below the synchrotron cooling time. At present, the shape of the electron cutoff, that could provide information on the particle acceleration/transport mechanism, is poorly constrained by the data, owing to the limited coverage in X rays restricted to energies $<10$~keV and to limited statistics in gamma rays (resulting from a combination of instrumental limitations and an intrinsic steep decline of the fluxes from the Klein-Nishina suppression of inverse-Compton emission). However, the datasets start constraining the level of turbulence in the cocoon magnetic field.

\acknowledgments
{\footnotesize
The support of the Namibian authorities and of the University of Namibia in facilitating the construction and operation of H.E.S.S. is gratefully acknowledged, as is the support by the German Ministry for Education and Research (BMBF), the Max Planck Society, the German Research Foundation (DFG), the Alexander von Humboldt Foundation, the Deutsche Forschungsgemeinschaft, the French Ministry for Research, the CNRS-IN2P3 and the Astroparticle Interdisciplinary Programme of the CNRS, the U.K. Science and Technology Facilities Council (STFC), the IPNP of the Charles University, the Czech Science Foundation, the Polish National Science Centre, the South African Department of Science and Technology and National Research Foundation, the University of Namibia, the National Commission on Research, Science \& Technology of Namibia (NCRST), the Innsbruck University, the Austrian Science Fund (FWF), and the Austrian Federal Ministry for Science, Research and Economy, the University of Adelaide and the Australian Research Council, the Japan Society for the Promotion of Science and by the University of Amsterdam.
We appreciate the excellent work of the technical support staff in Berlin, Durham, Hamburg, Heidelberg, Palaiseau, Paris, Saclay, and in Namibia in the construction and operation of the equipment. This work benefited from services provided by the H.E.S.S. Virtual Organisation, supported by the national resource providers of the EGI Federation.
This research has made use of data obtained from the \suz satellite, a collaborative mission between the space agencies of Japan (JAXA) and the USA (NASA). 
}

\bibliographystyle{JHEP}
{\footnotesize
\bibliography{ref}  
}

\end{document}